# Low Overhead Weighted-Graph-Coloring-Based Two-Layer Precoding for FDD Massive MIMO Systems

Abdelrahman A.H. Anis*, Bassant Abdelhamid, *Member, IEEE,* and Salwa Elramly, *Senior Member, IEEE*
Electronics and Electrical Communication Engineering Department
Faculty of Engineering, Ain Shams University
Cairo 11517, Egypt
{abdelrahman.anis, bassant.abdelhamid, salwa_elramly}@eng.asu.edu.eg

*Abstract*—A massive multiple-input multiple-output (MIMO) system, operating in Frequency Division Duplexing (FDD) mode of operation, suffers from prohibitively high overhead associated with downlink channel state information (CSI) acquisition and downlink precoding, due to the lack of uplink/downlink channel reciprocity. In this paper, a heuristic edge-weighted vertex-coloring based pattern division (EWVC-PD) scheme is proposed to alleviate the overhead of a two-layer precoding approach, in a practical scenario where the user clusters undergo serious angular-spreading-range (ASR) overlapping. Specifically, under a constraint of limited number of subchannels, an undirected edge-weighted graph (EWG) is firstly constructed, to depict the potential ASR overlapping relationship among clusters. Then, inspired by classical graph coloring algorithms, we develop the EWVC-PD scheme which mitigate the ASR by subchannel orthogonalization between clusters possessing serious ASR overlapping, and multiplexing the ones having slight ASR overlapping. Simulation results reveal that our scheme efficiently outperforms the existing pattern division schemes.

## I. INTRODUCTION

The unprecedented increase in mobile traffic and user terminals (UTs) need to be approached in future 5G cellular systems. A proposed solution to this challenge is to employ massive MIMO techniques, which has been widely investigated for almost a decade due to their potential significant improvement in spectral efficiency and energy efficiency by dense and adaptive spatial reuse [1]–[4]. However, these systems suffer from prohibitively high computational time complexity bearing on the precoding process. Time division duplex (TDD) mode can exploit the uplink/downlink channel reciprocity for estimating downlink CSI. Conversely, FDD mode is more advantageous than TDD one for the ease of backward compatibility with current cellular systems. Nonetheless, on the account of lack of channel reciprocity in FDD mode of operation, most approaches in FDD massive MIMO systems anticipate some sort of downlink training at the UTs with the aid of an information feedback from the UTs to the base station (BS) for downlink CSI acquisition. In particular, for massive number of antennas at the BS, the downlink FDD CSI acquisition leads to both significant bottleneck at the UTs and an unacceptable high signaling overhead for the uplink. An inspiring two-layer Joint Spatial Division and Multiplexing (JSDM) was proposed, to reduce the CSI training and acquisition in FDD downlink massive MIMO systems [5]–[7]. UTs are grouped into clusters with similar covariance eigen spaces, and their channel second order statistics is needed for the design of a pre-beamforming matrix, followed by a classical multiuser MIMO precoder that depends on the originating reduced-dimension channel matrix.

Several two-layer precoding schemes have been proposed in the literature [5], [7]–[11]. In [5], it was assumed that the UTs in different clusters have non-overlapping ASRs. However, there exist in practice some situations where such angular separation is not possible. Such situations are considered in [7]–[11]. In [8]–[10], clusters with slight ASR overlapping are scheduled on orthogonal subchannels, although they could be multiplexed on the same subchannel with increased spatial multiplexing gain. In [7], two pattern assignment algorithms were proposed: the first aims to maximize the orthogonalization between clusters along the different subchannels, while the second aims to maximize the multiplexing between clusters on the same subchannel. However, clusters with overlapped ASR are scheduled into orthogonal sub-channels with no constraint on the number of subchannels needed to achieve optimality. On the other hand, in [11], the subchannel number constraint is imposed, by allowing clusters with slight ASR overlapping to be multiplexed on the same subchannel. To the best of our knowledge, scheduling users with slight ASR overlapping on the same subchannel in order to limit the number of occupied subchannels, is analyzed only in [11]. The authors in [11] proposed a sub-optimal pattern assignment solution, using a graph theory-based pattern division (GT-PD) scheme, with low complexity, low signaling overhead.

To achieve near-optimal performance with consideration of the required computational complexity, we extend the foundation of [11] by further decreasing the computational complexity than the one needed by GT-PD scheme, by means of the proposed EWVC-PD scheme. In addition, a more articulated metric is introduced for pre-beamforming design that leads to under-utilization of patterns, i.e. more efficient use of the costly spectrum resources.

*Notation:* We use boldface lowercase (uppercase) letters for vectors (matrices), lowercase letters for scalars, and calligraphic letters for sets. $(X)^H$ denotes the Hermitian transpose, $rank\{X\}$ and $dim\{X\}$ denote the rank and the dimension of the matrix $X$. The $n \times n$ identity matrix is denoted by $I_n$. $x \sim \mathcal{CN}(\mu, \Sigma)$ indicates that $x$ is a complex circularly-symmetric Gaussian vector with mean $\mu$ and covariance matrix $\Sigma$. $Span\{X\}$ denotes the column subspace of $X$, whereas $Span^\perp\{X\}$ is the orthogonal complement of $Span\{X\}$, i.e. the left null space of $X$. The union, intersection and difference between two sets $\mathcal{X}$ and $\mathcal{Y}$ are respectively denoted by $\mathcal{X} \cup \mathcal{Y}$, $\mathcal{X} \cap \mathcal{Y}$ and $\mathcal{X} \setminus \mathcal{Y}$. $|\mathcal{X}|$ indicates the cardinality of the discrete set $\mathcal{X}$.

## II. SYSTEM AND CHANNEL MODELS

Consider a single cell multi-user massive MIMO downlink system operating in FDD mode. The cell is 120° sectorized and consists of one BS at the geometric center position. For each sector, the BS is equipped with a Uniform Linear Array (ULA) of $M$ ($\gg 1$) antennas serving $K$ active single-antenna User Terminals (UTs) per sector, as depicted in Fig. 1 (a).

Assuming Orthogonal Frequency Division Multiplexing (OFDM), we focus on a wide-sense stationary block fading wide band channel model [12], operating in a rich multi-path scattering environment with no line-of-sight propagation, at the sub-6 GHz band. Each flat fading sub-channel is referred to as "pattern". The dimensional channel matrix $H \in \mathbb{C}^{M \times K}$ is assumed to be fixed over the channel coherence block length of T channel uses. Suppose that the UTs are grouped into $G$ almost co-located user clusters with similar covariances by means of a user grouping algorithm [6], $K_g$ denotes the number of UTs in cluster $g$. The clusters are further separated into few pre-defined number of disjoint and mutually exclusive sets, previously referred to as "patterns", such that the clusters in the same pattern are served by the same flat fading sub-channel, whereas clusters in different patterns are served by different orthogonal sub-channels. This partitioning is essential in order to control the inter-cluster interference (ICI). The channel vectors are random processes with almost time-invariant second-order statistics [5], i.e. the channel covariance matrix for each user cluster is constant over a time period much longer than the instantaneous channel coherence time, which enables its efficient estimation, as analyzed in [13]–[15]. In a typical scattering scenario, even if a UT changes its position by several meters, the channel second order statistics remain unchanged. Assume the BS has full knowledge of the channel covariance matrices for each user cluster and the effective channels for each UT. Similar to [5], [7]–[9], [11], the one-ring channel model is adopted. The one-ring channel model describes the case of a BS that is elevated away from scatterers and is communicating with co-located UTs surrounded by a ring of local scatterers.

Let $h_{pgk} \in \mathbb{C}^M$ denote the channel gain vector from the BS to the $k$-th UT of the $g$-th cluster scheduled on the $p$-th pattern. Define the pattern set $\mathcal{C}_p$ to gather the clusters scheduled on the $p$-th pattern. $h_{pgk} \sim \mathcal{CN}(\mathbf{0}, C_{pgk})$, where $C_{pgk} \in \mathbb{C}^{M \times M}$ is the Hermitian semi-definite channel covariance matrix. The cluster channel matrix is defined as $H_{pg} = [h_{pg1}, h_{pg2}, \dots, h_{pgK_g}] \in \mathbb{C}^{M \times K_g}$. Since the UTs sharing the same cluster, have i.i.d. channels, with approximately the same covariance matrix, the angular reciprocity in FDD systems can be exploited [16].

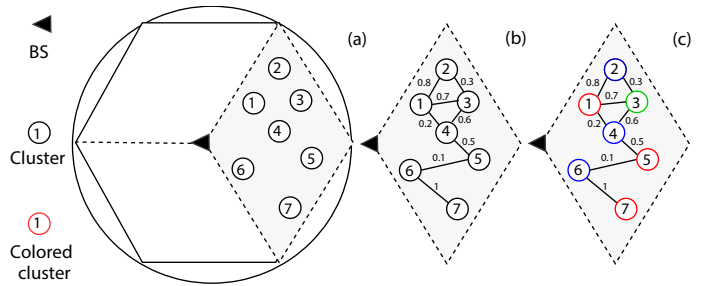

Fig. 1. Edge-weighted graph (EWG) construction and coloring. (a) Sectorized cell covering a number of UT clusters. (b) EWG construction. (c) Edge-weighted vertex-coloring based pattern division (EWVC-PD) scheme is applied.

Considering $r_g = rank\{C_g\}$, $\Lambda_g \in \mathbb{C}^{r_g \times r_g}$ is the diagonal matrix of the non-zero eigenvalues of $C_g$, $E_g \in \mathbb{C}^{M \times r_g}$ is the tall unitary matrix of the corresponding eigenvectors. From a wireless communication perspective, the eigenvectors indicate the directions on which signals are being transmitted, while the non-zero eigenvalues, $diag\{\Lambda_g\}$, signify the transmit powers allocated onto each direction. Using Karhunen-Loeve representation, the channel vectors $\{h_{pgk}\}$ can be written as in [5], [11]:

$$h_{pgk} = E_g \Lambda_g^{\frac{1}{2}} w_{pgk}, \quad (1)$$

where $w_{pgk}$ is a $r_g \times 1$ complex random vector with distribution $\sim \mathcal{CN}(\mathbf{0}, I_{r_g})$. Thanks to the directional structure of massive MIMO channels, the Toeplitz matrix $C_g$ can be asymptotically approximated by a circulant matrix [5], leading to the following eigenspace approximation sense where $E_g$ can be constructed from the $M \times M$ unitary Discrete Fourier Transform (DFT) matrix, whose elements are given by $[\mathbf{F}]_{l,n} = \frac{e^{-j2\pi l n / M}}{\sqrt{M}}$, as given in [5], [10], [11], where

$$E_g = [f_m : m \in \mathcal{J}_g], \quad (2)$$

is the DFT sub-matrix containing the columns with indices in the support set $\mathcal{J}_g$, which is defined as in [5]

$$\mathcal{J}_g = \left\{ m : \left(\frac{m}{M} - \frac{1}{2}\right) \in [-D\sin(\theta_g + \Delta_g), -D\sin(\theta_g - \Delta_g)], \right. \\ \left. m = 0, \dots, M - 1 \right\}, \quad (3)$$

where $D$ is the ratio of the spacing between the antenna elements relative to the wavelength $\lambda$, $\theta_g$ is the central azimuth angle of the local scattering ring of cluster $g$, and $\Delta_g$ is the angular spread, which can be approximated as $\Delta_g \cong arctan(s_g/d_g)$, with $s_g$ being the radius of the scattering ring, and $d_g$ being the distance between the BS and the ring center. $r_g$ equals the cardinality of the support set $\mathcal{J}_g$. Moreover, the support set $\mathcal{J}_g$ can be safely regarded as the ASR measure of cluster $g$. Also, the eigen value spectrum can be approximated as [5] by the function $S(\xi): \xi \in \left[\frac{-1}{2}, \frac{1}{2}\right] \to S(\xi) \in \mathbb{R}^+$, where

$$S(\xi) = \begin{cases} \dfrac{1}{2\Delta_g\sqrt{D^2 - \xi^2}} & \xi \in \mathcal{J}_g \\ 0 & \xi \notin \mathcal{J}_g \end{cases} \quad (4)$$

In addition, due to the small value of the angular spread $\Delta_g$, $\boldsymbol{R}_g$ possesses a low rank property, i.e. $r_g \ll M$, which facilitates meeting the tall unitary condition in [5]. Hence, the achievable rate of UTs with two-layer precoding is converged to the optimal single-UT rate, under condition of non-overlapped scattering rings. However, there exist in practice some situations where such angular separation is not possible. These situations are anticipated and mitigated in Section IV.

Since $G$ clusters are scheduled over $P$ patterns, let $\mathcal{C}_p$ denotes the set of clusters scheduled on the $p$-th pattern. The received signal at the UTs of cluster $g \in \mathcal{C}_p$ can be calculated as [5], [11]

$$\boldsymbol{y}_g = \boldsymbol{H}_{pg}^H \boldsymbol{U}_{pg} \boldsymbol{x}_g + \sum_{g' \in \mathcal{C}_p, g' \neq g} \boldsymbol{H}_{pg}^H \boldsymbol{U}_{pg'} \boldsymbol{x}_{g'} + \boldsymbol{z}_g, \quad (5)$$

where $\boldsymbol{U}_{pg}$ is $M \times V_g$ complex precoding matrix for cluster $g$, $\boldsymbol{x}_g$ is the $V_g \times 1$ vector of transmitted symbols from the BS to cluster $g$, such that $\mathbb{E}[\boldsymbol{x}_g \boldsymbol{x}_g^H] = \frac{\mathcal{P}_t}{K} \boldsymbol{I}_{V_g}$, with $\mathcal{P}_t$ indicating the average total transmitted power, $\boldsymbol{z}_g \sim \mathcal{CN}(\boldsymbol{0}, \sigma^2 \boldsymbol{I}_{K_g})$ is the additive complex white circularly Gaussian noise vector. The first term in Eq. (5) represents the desired signal, while the second and third terms represent the ICI and the uncorrelated noise, respectively.

## III. RELATED WORK

Under the JSDM transmission framework [5], the precoding matrix $\boldsymbol{U}_{pg}$ is designed as a product of two precoding matrices (layers):

$$\boldsymbol{U}_{pg} = \boldsymbol{U}_{pg}^{(1)} \boldsymbol{U}_{pg}^{(2)}, \quad (6)$$

where the first layer, the pre-beamformer, $\boldsymbol{U}_{pg}^{(1)} \in \mathbb{C}^{M \times N_g}$ is designed to eliminate the ICI, depending only on the channel second-order statistics, i.e. the channel covariance matrices. While the second layer $\boldsymbol{U}_{pg}^{(2)} \in \mathbb{C}^{N_g \times V_g}$ is designed to minimize the intra-cluster interference, based on the knowledge of the resulting reduced-dimensional channel matrix. This layering simplifies the precoding design problem, decouples the required channel statistics for each layer, and enables a successive dimensionality reduction, which in turn reduces the training and feedback overhead in FDD systems. Once the first layer is optimized, $\boldsymbol{U}_{pg}^{(1)}$ is used to construct the $N_g \times K_g$ effective reduced-dimension channel matrix $\overline{\boldsymbol{H}}_{pg}$ needed for designing $\boldsymbol{U}_{pg}^{(2)}$, where $\overline{\boldsymbol{H}}_{pg}^H = \boldsymbol{H}_{pg}^H \boldsymbol{U}_{pg}^{(1)}$. The dimension $N_g$ must satisfy the condition $V_g \leq N_g \leq r_g$. The more $N_g$ approaching near $r_g$, the more dominant directions available to cluster $g$, and the higher the spectral efficiency. Increasing $N_g$ is then considered within the first precoding layer design as will be shown in the following section.

Furthermore, with the aid of proper cluster assignment among patterns, it is possible to design the pre-beamforming matrix, as in [5], such that the ICI is completely or approximately eliminated as

$$\overline{\boldsymbol{H}}_{pg}^H = \boldsymbol{H}_{pg}^H \boldsymbol{U}_{pg'}^{(1)} \approx \boldsymbol{0}_{K_g \times N_g}, \forall g \neq g', \quad (7)$$

where the exact equality is satisfied with eigen-beamforming if $\text{Span}\{\boldsymbol{U}_{pg}^{(1)}\} \nsubseteq \text{Span}\{[\boldsymbol{U}_{pg'}^{(1)}: g' \neq g]\} \forall g$. Nonetheless, a condition is enforced on the number $K_g$ of UTs to be multiplexed per cluster, as given by [5]

$$dim\{\text{Span}\{\boldsymbol{U}_{pg}^{(1)}\} \cap \text{Span}^\perp\{[\boldsymbol{U}_{pg'}^{(1)}: g' \neq g]\}\} \geq K_g. \quad (8)$$

Correspondingly, the received signal in (5) can be rewritten as

$$\boldsymbol{y}_g = \overline{\boldsymbol{H}}_{pg}^H \boldsymbol{U}_{pg}^{(2)} \boldsymbol{x}_g + \boldsymbol{z}_g. \quad (9)$$

With the cancellation of the ICI, and by assuming that each BS has perfect effective CSIT, the achievable sum-rate of cluster $g$ is given by [4], [11]

$$\mathcal{R}_g = \log_2 det\{\boldsymbol{I}_{K_g} + (1/\sigma^2)\overline{\boldsymbol{H}}_{pg}^H \boldsymbol{U}_{pg}^{(2)} \boldsymbol{x}_g (\overline{\boldsymbol{H}}_{pg}^H \boldsymbol{U}_{pg}^{(2)} \boldsymbol{x}_g)^H\}. \quad (10)$$

We aim to maximize the system sum-rate by optimizing the first precoding layer, which can be tackled by solving the following optimization problem [11]:

$$(\mathbb{P}_1): \max_{\mathcal{C}_1, \mathcal{C}_2, \ldots, \mathcal{C}_P} \sum_{p=1}^{P} \sum_{g \in \mathcal{C}_p} \mathcal{R}_g$$

$$\text{subject to } \bigcup_{p=1}^{P} \mathcal{C}_p = \{1, 2, \ldots G\} \quad (11)$$

where the constraint ensures the complete solution, i.e. each cluster is assigned to one pattern and one pattern only. The optimization problem $\mathbb{P}_1$ can be solved by an exhaustive search algorithm (ESA), with a searching complexity of $O(P^G)$ over the different pattern assignments, which is considered prohibitively high for large values of $G$ and $P$. Moreover, ESA needs a training overhead to acquire the instantaneous effective channel knowledge for a pre-processing step, and a time-consuming two precoding layers calculation for each candidate pattern assignment, leading to infeasible time complexity relative to the limited channel coherence time. To address these challenges, a graph-theoretic heuristic algorithm is proposed in [11], where a sub-optimal solution is achieved, using a graph theory-based pattern division (GT-PD) scheme, with low complexity, low signaling overhead.

We extend the foundation of both [7], [11] by: (a) decreasing further the computational complexity than the one needed by GT-PD scheme in [11]; (b) introducing a more articulated metric for pre-beamforming design that leads to under-utilization of patterns, more efficient use of the spectrum resources; and (c) integrating both orthogonalization and multiplexing objectives, addressed in [7], into our proposed EWVC-PD scheme, in order to attain their mutual benefits, leading to an effective adjustable scheme pertaining to the SNR requirements of the propagation scenario under study.

## IV. PROPOSED EWVC-PD SCHEME

### A. Pre-beamforming Design

Our goal is to design the first layer precoders $U_{pg}^{(1)}$ such that condition (7) is satisfied for all clusters. Given a certain pattern assignment, we employ the DFT based eigen-beamforming based on the channel covariances of the selected clusters per pattern where $U_{pg}^{(1)}$ is designed by projecting the $span\{U_{pg}^{(1)}\}$ into $Span^\perp\{\Psi_{pg}\}$, where $\Psi_{pg}$ is constructed by $E_{g'}$ of all cluster $g'$ scheduled on the same pattern as cluster $g$, which can be calculated as [17]

$$\Psi_{pg} = \left[ f_m : m \in \bigcup_{g' \in \mathcal{C}_p, g' \neq g} \mathcal{J}_{g'} \right] \quad (12)$$

Since the DFT matrix is unitary, $U_{pg}^{(1)}$ can be written as [17]

$$U_{pg}^{(1)} = \left[ f_m : m \in \left( \mathcal{J}_g \setminus \bigcup_{g' \in \mathcal{C}_p, g' \neq g} \mathcal{J}_{g'} \right) \right]. \quad (13)$$

where the first layer precoder is designed with its unique non-overlapped set of supports, whose cardinality is equal to the dimension of the effective channel matrix, $N_g$, given as [11]

$$N_g = rank\{U_{pg}^{(1)}\} = \left| \mathcal{J}_g \setminus \bigcup_{g' \in \mathcal{C}_p, g' \neq g} \mathcal{J}_{g'} \right|. \quad (14)$$

For two clusters $g$ and $g'$, it can be deduced that [11]:

$N_g + N_{g'} = \gamma_{g,g'} - \eta_{g,g'}$,

where $\gamma_{g,g'} = r_g + r_{g'} - 2|\mathcal{J}_g \cap \mathcal{J}_{g'}|$,

$$\eta_{g,g'} = \left| \mathcal{J}_g \cap \left( \bigcup_{i \in \mathcal{C}_p, i \neq g, i \neq g'} \mathcal{J}_i \right) \right| + \left| \mathcal{J}_{g'} \cap \left( \bigcup_{j \in \mathcal{C}_p, j \neq g, j \neq g'} \mathcal{J}_j \right) \right| \quad (15)$$
$$- \left| \mathcal{J}_g \cap \mathcal{J}_{g'} \cap \left( \bigcup_{l \in \mathcal{C}_p, l \neq g, l \neq g'} \mathcal{J}_l \right) \right|.$$

Note that $\gamma_{g,g'}$ requires only the knowledge of the support sets $\mathcal{J}_g$ and $\mathcal{J}_{g'}$ only, whereas $\eta_{g,g'}$ depends on the support sets of all clusters scheduled on the same pattern as cluster $g$ and $g'$. Also, $\gamma_{g,g'}$ is a non-negative real number, and its inverse was chosen to represent the ASR metric in [11].

### B. Graph Construction

As stated previously in Eq. (3) and illustrated in Fig. 1 (b), the support set $\mathcal{J}_g$ can be regarded as our measure for ASR. To best formulate the ASR overlapping metric between clusters, consider the ASR overlapping between two clusters only, $g$ and $g'$. Hence, the ASR overlapping scenarios can be classified to three scenarios as follows.

- Non-overlapped ASR scenario: $\mathcal{J}_g \cap \mathcal{J}_{g'} = \emptyset$, i.e. the clusters do not share any common support. In such case, full "multiplexing" is employed, where both can be perfectly spatially multiplexed on the same pattern. The pre-beamforming matrices can then be calculated by eigen-beamforming over all available supports as $U_{pg}^{(1)} = E_g$, and $U_{pg'}^{(1)} = E_{g'}$.

- Fully overlapped ASR scenario: $\mathcal{J}_g \subseteq \mathcal{J}_{g'}$, or vice versa. Consequently, at least one of the clusters does not possess any non-overlapped spectral interval to transmit on, i.e. ICI cannot be recovered. In such case, "orthogonalization" is employed, where the clusters are served on different patterns $p$ and $p'$, and the pre-beamforming matrices are used to transmit over all the channel eigenmodes to each group separately as $U_{pg}^{(1)} = E_g$, and $U_{p'g'}^{(1)} = E_{g'}$.

- Partially overlapped ASR scenario: $\mathcal{J}_g \cap \mathcal{J}_{g'} \neq \emptyset, N_g \neq 0, N_{g'} \neq 0$. In this case, we choose to employ either multiplexing or orthogonalization based on our ASR overlapping metric $w_{g,g'}$ defined as

$$w_{g,g'} = \frac{2\epsilon |\mathcal{J}_g \cap \mathcal{J}_{g'}|}{r_g + r_{g'}}, w_{g,g'} \in [0,1], \quad (16)$$

where $\epsilon$ is a proportionality constant denoting the relative gain of orthogonalization over multiplexing. Same as $\gamma_{g,g'}$, $w_{g,g'}$ requires only the knowledge of the support sets $\mathcal{J}_g$ and $\mathcal{J}_{g'}$. The closer $w_{g,g'}$ to unity, the less number of unique non-overlapped available supports compared to the number of common supports left if multiplexing is employed. This indicates a more severe ASR, with reducing the system sum rate. The drawback of multiplexing in this case is that a significant portion of the transmit power is lost. Hence, orthogonalization is favored in this case. On the other hand, the closer $w_{g,g'}$ to zero, the more the multiplexing is favored over orthogonalization, as the transmit energy loss becomes more tolerable compared to the spatial multiplexing gain introduced when more users can be served simultaneously on the same pattern.

For addressing the pattern assignment problem, an undirected EWG $\mathcal{G} = \{\mathcal{V}, \mathcal{E}, w\}$ is constructed, for which each vertex in set $\mathcal{V}$ denotes a cluster $g$, with $G = |\mathcal{V}|$, each edge $\{g, g'\} \in \mathcal{E}, \forall g \neq g'$ is allocated a non-negative real weight equal to $w_{g,g'}: \mathcal{E} \to \mathbb{R}_0^+$. For simplification, the edges with negligible $w_{g,g'}$ are omitted. The degree of a vertex is equal to the sum of weighted-edges connecting the vertex to its neighbors and is denoted by $deg(g)$. $\Pi$ is a permutation sequence of vertices, sorted by the descending order of vertex degree, where $\Pi_i$ is the $i$-th element in $\Pi$. The EWG is characterized as $P^*$-colorable if the EWG graph can be colored by $P^*$ colors, where $P^* \leq P$. The optimization problem $\mathbb{P}_1$ can then be approached by $\mathbb{P}_2$ as an edge-weighted vertex-coloring problem. Given a fixed number of $P$ available patterns, our task is to identify a complete pattern assignment solution $\mathcal{C} = \{\mathcal{C}_1, \mathcal{C}_2, \dots, \mathcal{C}_P\}$ that minimizes the objective function $\mathbb{P}_2$ as

$$(\mathbb{P}_2): \min_{\mathcal{C}} f(\mathcal{C}) = \sum_{p=1}^{P} \sum_{\substack{g,g' \in \mathcal{C}_p, \\ g' \neq g}} w_{g,g'},$$

$$\text{subject to } \bigcup_{p=1}^{P} \mathcal{C}_p = \{1,2,\dots G\}, \quad (17)$$

where any solution for which $f(\mathcal{C}) = 0$, corresponds to a feasible proper $P$-colored solution. If for a certain pattern $p$, $f(\mathcal{C}_p) = 0$, then all vertices $g \in \mathcal{C}_p$ form an independent set, i.e. the vertices are not connected by any edges. The combinatorial optimization problem $\mathbb{P}_2$ depends only on the channel time-invariant second-order statistics, which alleviates the need for both effective CSI at the BS and sum-rate calculation for every candidate solution. We can also see that $\mathbb{P}_2$ is a permutation-based optimization, whose optimal solution can also be obtained using the ESA as with $\mathbb{P}_1$. Given the high computational complexity for performing ESA, a heuristic edge-weighted vertex-coloring based pattern division (EWVC-PD) scheme is proposed as follows.

*C. Graph Coloring*

The conventional graph coloring problem (GCP) is a classic problem [18], where all vertices of a graph have to be colored in such a way that (a) no vertices joined by an edge are given the same color, and (b) the number of distinct colors used is minimized, as illustrated in Fig. 1 (c). Many efficient algorithms have been proposed in the literature [19], [20]. However, for our considered application, the second criterion of GCP should be replaced by another criterion maintaining $P$, the number of colors available, as a constraint, whereas the sum of edge weights in $\mathbb{P}_2$ is the one being minimized. To address these challenges, the EWVC-PD algorithm is proposed to reduce the ICI under the constraint of limited patterns. Inspired by the unweighted-edge Largest Degree Ordering (LDO) algorithm [19], the vertices are sorted according to their degrees in descending order and colored in a sequential way with reused colors as best as possible. The proposed EWVC-PD scheme greedily assigns different patterns to clusters having a large weighted-edge. However, unlike the LDO algorithm, the EWVC-PD scheme allows the possibility for two clusters, having small ASR overlapping $w_{g,g'}$, to be multiplexed on the same pattern. The proposed EWVC-PD scheme is summarized in Algorithm 1, which is mainly comprised of the following three parts:

*1) Initialization (Steps 2-4):* Firstly, two clusters $g_1, g_2$ with the largest weighted-edge are selected in step 3, then assigned to distinct patterns $\mathcal{C}_1, \mathcal{C}_2$ in steps 3 and 4. The set $\Omega$ of uncolored clusters and the permutation sequence $\Pi$ are both updated by deleting the colored clusters $g_1, g_2$.

*2) Phase I (Steps 6-16):* This phase consists of initially selecting the uncolored clusters $\in \Pi$, forming independent sets and assigning them to patterns, till the first cluster that cannot join any independent set formed in any color, the algorithm then proceeds to Phase II. Interestingly, Phase I is an unweighted-edge graph coloring algorithm, which exploits the common case of sparse graphs scenarios that can be colored with $f(\mathcal{C}) = 0$ without violating the color constraint, in which case all clusters are colored by Phase I algorithm, and Phase II is bypassed. Hence, the complexity is greatly reduced to $O(G^2)$.

*3) Phase II (Steps 18-23):* This phase is the main algorithm employing weighted-edge graph coloring under color constraint. The cluster selection is based on the order of minimizing $f(\mathcal{C})$, by means of a priority parameter $\delta_{pg}$, which is defined in step 19 as the weight sum of the edges introduced by cluster $g$, if the cluster $g$ is assigned to pattern $p$. Then, the cluster $g_0$ with the largest potential ICI out of the uncolored set $\Omega$ will be selected in step 20, whose pattern assignment should be considered preferentially. The cluster assignment in step 21 aims to schedule the selected cluster $g_0$ to the pattern $p_0$ having the smallest ICI introduced by the addition of cluster $g_0$. Finally, the cluster $g_0$ is added to the pattern set $\mathcal{C}_{p_0}$ in step 22. This loop will be carried out in a sequential way until all clusters are scheduled with their corresponding patterns. The computational complexity of this phase is $O(PG^2)$.

---

**Algorithm 1:** Proposed EWVC-PD Algorithm
---
**Input:**
  Number of patterns $P$, and permutation sequence $\Pi$;
  The constructed EWG: $\mathcal{G} = \{\mathcal{V}, \mathcal{E}, w\}$.
**Output:**
  Pattern assignment $\mathcal{C} = \{\mathcal{C}_1, \mathcal{C}_2, \dots, \mathcal{C}_P\}$.
1: **Initialization:**
2: $\mathcal{C}_p \leftarrow \emptyset, \forall p = 1, 2, \dots P$.
3: $\langle g_1, g_2 \rangle \leftarrow \arg\max\limits_{g \neq g'}\{w_{g,g'}\}, \forall g, g' \in \mathcal{V}$
4: $\mathcal{C}_1 = \{g_1\}, \mathcal{C}_2 = \{g_2\}, \Omega \leftarrow \mathcal{V} \setminus \{g_1, g_2\}, \Pi \leftarrow \Pi \setminus \{g_1, g_2\}$
5: **Phase I:**
6: **for** $i = 1$ to $|\Pi|$ **do**
7:   **for** $p = 1$ to $P$ **do**
8:     **if** $(\mathcal{C}_p \cup \{\Pi_i\})$ is an independent set **then**
9:       $\mathcal{C}_p \leftarrow \mathcal{C}_p \cup \{\Pi_i\}, \Omega \leftarrow \Omega \setminus \{\Pi_i\}$
10:       **break**
11:     **end**
12:   **end**
13:   **if** $\Pi_i$ is still uncolored **then**
14:     **break**
15:   **end**
16: **end**
17: **Phase II:**
18: **while** $\Omega \neq \emptyset$ **do**
19:   $\delta_{pg} \leftarrow \sum_{g' \in \mathcal{C}_p, g' \neq g} w_{g,g'}, \forall g \in \Omega, \forall p$
20:   $g_0 \leftarrow \arg\max\limits_{g \in \Omega}\{\max\limits_p\{\delta_{pg}\}\}$
21:   $p_0 \leftarrow \arg\min\limits_p\{\delta_{pg_0}\}$
22:   $\mathcal{C}_{p_0} \leftarrow \mathcal{C}_{p_0} \cup \{g_0\}, \Omega \leftarrow \Omega \setminus \{g_0\}$
23: **end**
24: **return** $\mathcal{C} = \{\mathcal{C}_1, \mathcal{C}_2, \dots, \mathcal{C}_P\}$.

---

The proposed EWVC-PD scheme is able to mitigate ICI and increase the sum-rate under the constraint of limited patterns, while depending only on the channel covariance matrices without the need for effective CSIT knowledge. Furthermore, the computational complexity of the proposed algorithm is $O(G^2)$ at best scenarios of sparse graphs, and $O(PG^2)$ at worst ones, which is much smaller than the computational complexity of the ESA $O(P^G)$.

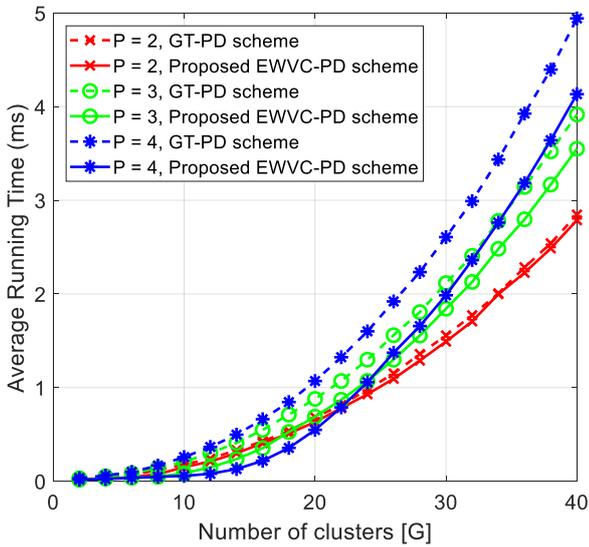

Fig. 2. Average CPU running time versus the number of clusters $G$, for the MATLAB code of the GT-PD scheme and the proposed EWVC-PD scheme, for varying number of patterns $P$ at $M = 128$.

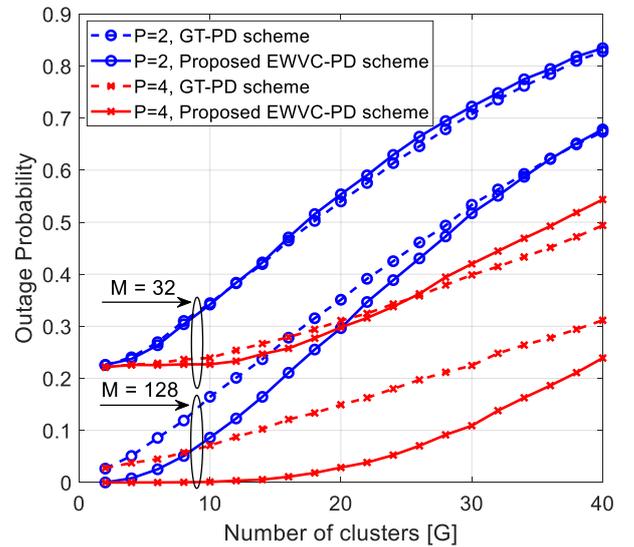

Fig. 3. Outage probability versus the number of clusters $G$, for the GT-PD scheme and the proposed EWVC-PD scheme, for varying number of patterns $P$ at $M = 32, 128$.

TABLE I.    SIMULATION PARAMETERS

| | |
|---|---|
| Number of BS antennas per sector $M$ | 32, 128 |
| Number of clusters per sector $G$ | $2 \leq G \leq 40$ |
| Number of available patterns $P$ | 2,3,4 |
| Number of users per cluster $K_g$ | 2 |
| Number of data symbols per cluster $V_g$ | 2 |
| Cell Radius $d_{max}$ | 600 m |
| Average transmit power $\mathcal{P}_t$ | 10 dB |
| Noise Variance $\sigma^2$ | 1 |
| Operating frequency $f$ | 2 GHz |
| ULA antenna spacing $\lambda D$ | $\lambda/2$ |

V. SIMULATION RESULTS

In this section, we investigate the performance of the proposed EWVC-PD scheme through Monte-Carlo simulations. The system parameters are summarized in TABLE I. Clusters are randomly and uniformly distributed in the cell sector. The one-ring channel vectors are generated from Eq. (1). For simplicity, the zero forcing beamforming is adopted for the design of the second layer precoder $\boldsymbol{U}_{pg}^{(2)}$ [21]. The algorithms were executed on MATLAB R2017b, on an Intel Core i7 2.00 GHz processor and 6 GB DDR3 RAM running Windows 10 Enterprise. Simulations used floating-point arithmetic.

The average CPU running time is depicted in Fig. 2. It considers the proposed EWVC-PD compared with the GT-PD scheme in [11]. To fairly compare the running time of these graph coloring algorithms, we used the same ASR overlapping metric in Eq. (16). We see that the EWVC-PD outperforms at high values of $G$, at $P = 3$ and 4, while achieving almost the same CPU time at $P = 2$. The reason is, at relaxed $P$ constraint, the Phase I initialization algorithm encompasses more clusters than the ones scheduled by Phase II. Whereas Phase II dominates scheduling more clusters at tight $P$ constraint, whose time approximates to the GT-PD scheme.

Moreover, Phase II algorithm is almost bypassed at relatively lower values of $G$ with relaxed $P$ constraint, which can lead to a significant time reduction, of approximately 40% reduction at $G = 14$ and $P = 4$.

Fig. 3 shows more directly the relationship between the outage probability and the number of BS antennas $M$. We can see that choosing to increase $M$, considerably alleviates the outage probability by EWVC-PD scheme compared to GT-PD scheme, as opposed to lower values of $M$. In addition, the outage probability decreases significantly compared to that of GT-PD scheme, at relaxed $P$ constraint and increased $M$, of approximately one order of magnitude reduction at $G = 16$, $M = 128$, and $P = 4$.

VI. CONCLUSION

In this paper, we have proposed a pattern division scheme based on weighted-edge vertex-coloring to ameliorate inter-cluster interference in a two-layer precoding approach. The proposed EWVC-PD scheme has several advantages in massive MIMO systems: (a) There is no need to acquire the instantaneous effective CSI at the BS for every candidate pattern division solution; (b) the two phases structure of the algorithm reduces the computational time complexity significantly compared to the existing GT-PD algorithm; (c) the articulated ASR metric efficiently limit the use of the costly spectrum resources; and (d) the results showed that EWVC-PD scheme achieves clear gains, in terms of outage probability, over the GT-PD scheme, with the increased scaling of the number of BS antennas. For future work, it would be interesting to integrate the implementation of two-layer precoding, jointly with efficient downlink CSI training and uplink information feedback.